\documentclass[11pt]{article}
\usepackage{graphics}
\usepackage{amsfonts}
\usepackage{amssymb}
\usepackage{amsbsy}
\def\half{{\scriptstyle\frac{1}{2}}}
\def\crop{\prod^{\circlearrowright}}
\def\sign{\mathrm{sign}}
\def\i{\mathrm{i}}
\def\bphi{\boldsymbol{\varphi}}

\def\H{{\mathcal H}}
\def\Ul{\mathfrak{Q}}

\def\Fl{\mathfrak{F}}
\def\Uc{\mathsf{Q}}
\def\Ucn{\mathsf{Q}_{\mathrm{naive}}}
\def\C{\mathsf{C}}
\def\Fc{\mathsf{F}}
\def\Sc{\mathsf{S}}
\def\e{e}
\def\x{\mathfrak{z}}
\def\qe{\epsilon}
\def\f{f}
\def\qem{\boldsymbol{\epsilon}_q}
\def\qed{\boldsymbol{\epsilon}_{q^{\theta^2}}}

\def\a{\mathsf{a}}
\def\b{\mathsf{b}}
\def\v{\mathfrak{v}}
\def\w{\mathfrak{w}}
\def\vc{\mathsf{v}}
\def\wc{\mathsf{w}}

\def\R{\mathsf{R}}

\def\sla{{(\lambda)}}
\def\smu{{(\mu)}}
\def\bka{{\scriptstyle{(\kappa)}}}
\def\bla{{\scriptstyle{(\lambda)}}}
\def\bmu{{\scriptstyle{(\mu)}}}
\def\bnu{{\scriptstyle{(\nu)}}}
\def\binf{{\scriptstyle{(\infty)}}}
\def\blamu{{\scriptstyle{(\lambda,\mu)}}}

\def\bea{\begin{eqnarray*}\\}
\def\eea{\\\end{eqnarray*}}

\begin{document}
\renewcommand{\thefootnote}{\fnsymbol{footnote}}

\begin{center}{\Large\bf Algebraic Quantization of
Integrable Models in Discrete Space-time}
\bigskip

{\large\bf L.D.\ Faddeev and A.Yu.\ Volkov\footnote
{Supported by Russian Foundation of Fundamental Research
and Finnish Academy.}}

{St. Petersburg Steklov Mathematical Institute
and Helsinki Institute of Physics}
\end{center}
\bigskip
\begin{quote}{\bf ABSTRACT.} Just like decent classical
difference-difference systems define symplectic maps
on suitable phase spaces, their counterparts with
properly ordered noncommutative entries come as
Heisenberg equations of motion for corresponding quantum
discrete-discrete models. We observe how this idea applies
to a difference-difference counterpart of the Liouville
equation. We produce explicit forms of of its evolution
operator for the two natural space-time coordinate systems.
We discover that discrete-discrete models inherit crucial
features of their continuous-time parents like locality
and integrability while the new-found algebraic
transparency promises a useful progress in some branches of
Quantum Inverse Scattering Method.
\end{quote}
\bigskip

\section{Introduction}

In this paper we intend to elucidate the
algebraic part of our approach to the
quantum integrable models in 1+1 dimensional
discrete space-time, developed during last five
years. We shall not give a complete survey of
our publications
(Faddeev and Volkov 1993,1994; Faddeev 1994; Volkov 1997a,b)
because it would take too much space. We believe,
that the algebraic side is most instructive and
original; more analytic side will be mentioned
only briefly with references to (Faddeev 1994)

Discrete space-time models (DSTM) in soliton
theory have acquired a prominent role from the very
advent of this part of mathematical physics. The
first examples of such models were proposed by
Hirota (1977)
as discrete analogues of the major continuous
soliton models.
Subsequent development was carried on mostly by Dutch
group, see (Nijhoff and Capel 1995) and references therein.
The recent resurgence of interest towards DSTM is
connected with several new ideas:

1. The nonlinear equations for the family of
transfer matrices
$ T_{S}(\lambda) $
in the framework of the Thermodynamic Bethe Ansatz
can be considered as DSTM with spin
$ S $
and rapidity
$ \lambda $
being discrete variables (Kl\"umper and Pearce 1992;
Kuniba {\em et al.}1994; Krichever {\em et al.}1996).
Of course, the rapidity assumes the continuous values,
but only discrete shift $\lambda \to q \lambda $
enters the equations.

2. A.~Bobenko and U.~Pinkal with collaborators
develop the discrete analogue of classical
continuous 2-dimensional differential geometry,
see (Bobenko and Pinkal 199?) and references therein.

3. Quantum version of DSTM revealed a new type of
symmetry, giving the discrete analogue of current
algebra and Virasoro algebra (Faddeev and Volkov 1993;
Volkov 1997c).
Moreover, quantum DSTM seems to be rather universal,
giving both massless (conformal) and massive models
in continuous limit.

For the evident methodological reason we shall
illustrate our approach on a concrete example.
For that we have chosen the most prominent model
of physics and geometry --- the Liouville model.
More involved Sine-Gordon model will be touched
upon only briefly. Incidentally, the latter was
already a subject of our earlier publications.
We shall not discuss the usual paraphernalia of integrable
models such as zero-curvature representation, Lax
equation and Bethe ansatz. We shall simply present
the main dynamical object --- the evolution operator,
realizing the elementary time-shift. Its natural place
inside the Algebraic Bethe Ansatz is discussed in recent
lectures of one of authors (Faddeev 1996).
We believe, that our explicit formulas are
interesting enough as they stand so we want to present
them in its clearest form, independent of original
derivation.

We begin with the reminder of the classical Liouville
model and its Hamiltonian interpretation. This will
play the role of the starting point for the subsequent
deformations: discretization of space on which
hamiltonian data are given, and quantization. As a
result we shall get a suitable algebra of observables.
Finally the time evolution will be defined in terms of
a certain automorphism of this algebra. The discrete
time equations of motion produced by this automorphism
will be shown to be a natural analogue of the
corresponding classical equations. The integrability
of the model will be confirmed by presenting an explicit
set of conservation laws.

\section{Classical differential equation}

As the goals declared in the Introduction suggest, this
time we shall consider the evergreen Liouville's
Equation (LE)
\[ \frac{\partial^2\varphi}{\partial t^2}
   -\frac{\partial^2\varphi}{\partial x^2}
   =e^{-2\varphi}                                         \]
a Hamiltonian 1+1-dimensional field theory with $x$ denoting
the spatial coordinate and $t$ serving as time.
The Cauchy data
\[ \varphi(x,t)|_{_{t=0}}=\varphi(x)\qquad\qquad
   \frac{\partial\varphi}{\partial t}(x,t)|_{_{t=0}}
   =\varpi(x)                                             \]
can be equipped with the canonical Poisson bracket
\[ \{\varpi(x),\varphi(y)\}=\delta(x-y)\qquad\qquad
   \{\varpi(x),\varpi(y)\}=\{\varphi(x),\varphi(y)\}=0    \]
so that the evolution goes the Hamiltonian way in its
most familiar
\[ \dot{\varphi}=\{H,\varphi\}=\varpi\qquad\qquad
   \dot{\varpi}=\{H,\varpi\}=\varphi''+e^{-2\varphi},     \]
the Hamiltonian being
\[ H=\half\int dx\;(\varpi^2+(\varphi')^2+e^{-2\varphi}). \]
The periodic boundary conditions
\[ \varphi(x+2\pi)=\varphi(x)\qquad\qquad
   \varpi(x+2\pi)=\varpi(x)                               \]
pose no problem provided the Poisson bracket employs a
2$\pi$-periodic delta-function rather than the ordinary one.

Of the equation's specific features the Liouville's formula
\[ \e^{-2\varphi(x,t)}
   =\frac{f'(\xi)g'(\tau)}{(f(\xi)-g(\tau))^2}            \]
\[ x=\xi-\tau\qquad\qquad t=\xi+\tau                      \]
making a solution out of two arbitrary functions, is the
ultimate. It is there, according to (Gervais and Neveu 1982;
Faddeev and Takhtajan 1985), where the real
Hamiltonian theory of LE begins. We shall not reach that
high in this paper.

\section{Classical difference equation}

The best lattice approximation to LE
\[ \e^{\varphi(x,t-\Delta)}\e^{\varphi(x,t+\Delta)}
   -\e^{\varphi(x-\Delta,t)}\e^{\varphi(x+\Delta,t)}
   =\Delta^2                                              \]
is due to R. Hirota (1987) like virtually every decent
difference-difference equation. In order to make its
transformation into LE under limit $\Delta\rightarrow 0$
more obvious one may recompose it like this:
\[ \sinh\half\bigg(\varphi(x,t-\Delta)+\varphi(x,t+\Delta)
   -\varphi(x-\Delta,t)-\varphi(x+\Delta,t)\bigg)         \]
\[ =\Delta^2
   \e^{-\half(\varphi(x,t-\Delta)+\varphi(x,t+\Delta)
   +\varphi(x-\Delta,t)+\varphi(x+\Delta,t))}   .         \]
Now as the mission of the lattice spacing $\Delta$ is over,
it is only natural to have everything suitably rescaled
\[ (x,t)\longrightarrow(\Delta x,\Delta t)                \]
\[ \e^\varphi\longrightarrow \Delta\e^\varphi             \]
or just set $\Delta$=1. Either way, the
Difference Liouville Equation (DLE) takes its final form
\[ \e^{\varphi_{j,k+1}}\e^{\varphi_{j,k-1}}
   -\e^{\varphi_{j+1,k}}\e^{\varphi_{j-1,k}}=1            \]
\[ j+k\equiv1\pmod2                                       \]
where the change for subscripts manifests that 
the `space-time' is now a ${\mathbb{Z}\,}^{2}$ lattice while
the second line specifies which half of that lattice the
equation will actually occupy. This half itself makes a
square lattice turned by fourty five
degrees with respect to the original one and twice less
dense. The values of $\varphi$ on a `saw' formed by vertices
with $k$ equal either 0 or 1
\[ \varphi_{j,0}=\varphi_{j}\qquad\mbox{for even $j$}     \]
\[ \varphi_{j,1}=\varphi_{j}\qquad\mbox{for odd $j$}      \]
make a reasonable Cauchy data, that is they are just
sufficient to have the whole system resolved step by step.
This has everything to do with the second-order nature of
the original continuum equation whose Cauchy data
combine the present and a little bit of the future
represented by $\varphi(x)$ and $\varpi(x)$ respectively.

Quite expectedly, there exists a Poisson bracket preserved
under evolution along $k$-direction governed by DLE.
However, it turns out more complicated than one might have
wished a lattice deformation of the canonical one would be:
\[ \{\varphi_i,\varphi_j\}=\varsigma(i,j)                 \]
with
\[ \varsigma(i,j)=\left\{\begin{array}{l}
   0\qquad\qquad\mbox{if $i-j\equiv 1\pmod2$}\\\\
   (-1)^{\half(i+j+1)}\sign(i-j)\qquad
   \mbox{otherwise}\end{array}\right.                     \]
Such is the price for the ultimate simplicity of the
equation. This would be too much if we had lost the
option of periodic boundary condition
\[ \varphi_{j+L}=\varphi_j.                               \]
Fortunately, we had not. If the period is chosen properly
\[ L=2M\qquad\qquad M\equiv1\pmod2                        \]
the bracket remains intact provided the above
description of $\varsigma$ applies when $|i-j|\leq L$
and extends periodically
\[ \varsigma(i\pm L,j)=\varsigma(i,j)                     \]
elsewhere. Those still insisting on an easier bracket 
can change the variables
\[ \phi_j=\half(\varphi_{j+1}+\varphi_{j-1})              \]
and have it
\bea && \{\phi_i,\phi_j\}=0\qquad\mbox{if $|i-j|\neq 1$}\\\\
   && \{\phi_{j-1},\phi_j\}=\frac{(-1)^j}{2}            \eea
at the expence of a busier equation
\[ \e^{2\phi_{j,k+1}}\e^{2\phi_{j,k-1}}
   =(1+\e^{2\phi_{j+1,k}})(1+\e^{2\phi_{j-1,k}})          \]
\[ j+k\equiv0\pmod2.                                      \]
Either way, the prospect of
dealing with discrete Poisson maps is hardly encouraging.
That is why we choose to leave the classical equations alone
and go quantum. Before we do, let us round out the classical
part with a beautiful, if irrelevant for our current
agenda, discrete Liouville formula:
\[ \e^{-2\phi_{j,k}}=\e^{-\varphi_{j+1,k}-\varphi_{j-1,k}}
   =\frac{(f_{m+1}-f_{m})(g_{n+1}-g_{n})}
   {(f_{m+1}-g_{n})(f_{m}-g_{n+1})}                       \]
\[ j=m-n\qquad\qquad k=m+n+1     .                        \]

\section{Algebra of observables}

One dilemma about quantization is whether
to develop it in terms of the bare $\varphi$'s or stick to
the variables actually entering the equation, that is
the exponents
\[ v_j=\e^{\varphi_j}\qquad\qquad
   \{v_i,v_j\}=\varsigma(i,j)v_i v_j                      \]
The respective Heisenberg- and Weyl-style quantum
algebras are
\[ [\bphi_i,\bphi_j]=\i\hbar\gamma\varsigma(i,j)          \]
\[ \bphi_{j+L}=\bphi_j                                    \]
with the usual lot in r.h.s. comprising the imaginary
unit $\i$, the Plank constant $\hbar$ and the coupling
constant $\gamma$; and
\[ \v_i\v_j=q^{\varsigma(i,j)}\v_j\v_i                    \]
\[ \v_{j+L}=\v_j                                          \]
with all packed in a single $q$uantisation constant
\[ q=\e^{\i\hbar\gamma}.                                  \]
Weyl-style algebra may be viewed as a subalgebra of the
Heisenberg-one
\[ \v_j=\e^{\bphi_j}                                      \]
but not the other way round. Roughly speaking,
the latter also accomodates non-integer powers
\[ \v^\alpha_j=\e^{\alpha\bphi_j}                         \]
not allowed in the former.

Another dilemma is whether to place $q$ on the unit circle
or not. The first option is obviously involution-friendly.
It allows for both unitary
\[ \v_i^*=-\v^{-1}_i                                      \]
and selfadjoint
\[ \v_i^*=\v^{}_i                                         \]
pictures, the former offering the luxury of dealing with
bounded operators if at the expence of complications
in representation theory due to the arithmetics of $q$
while the latter actually being the one relevant for the
true Liouville model. On the other hand, $q$ inside
(or outside) the circle is favoured in q-algebra but
whether it is good for something else remains to be seen.

We choose not to take sides before time and conclude the
Section on a more practical note.
Let us introduce, for future use, quantum counterparts of
the $\e^{2\phi}$-variables
\[ \w_j=\v_{j+1}\v_{j-1}=\v_{j-1}\v_{j+1}                 \]
and compile a list of emerging commutation relations
\bea && \v_j\w_j=q^{2(-1)^j}\w_j\v_j                    \\\\
   && \w_{j-1}\w_j=q^{2(-1)^j}\w_j\w_{j-1}              \\\\
   && \v_i\w_j=\w_j\v_i\qquad
   \mbox{if $i\neq j\pmod{L}$}                          \\\\
   && \w_i\w_j=\w_j\w_i\qquad
   \mbox{if $|i-j|\neq 1\pmod{L}$}.                     \eea

\section{Evolution operator}

Given an invertible operator $\Ul$, one can make the algebra
of observables `evolve'
\[ \cdots\longmapsto\Ul\x\Ul^{-1}\longmapsto\x\longmapsto
   \Ul^{-1}\x\Ul\longmapsto
   \Ul^{-2}\x\Ul^{2}\longmapsto\cdots                     \]
hoping that the evolution of generators
\[ \v_{j,k+2}=\Ul^{-1}\v_{j,k}\Ul\qquad j+k\equiv0\pmod2  \]
\[ \v_{2a,0}=\v_{2a}\qquad\qquad\quad\v_{2a-1,1}=\v_{2a-1}\]
manages to solve some nice and local equations,
for instance,
\[ \v_{j,k+1}\v_{j,k-1}-q^{-1}\v_{j-1,k}\v_{j+1,k}=1.     \]
As a matter of fact, this is exactly what happens if
\[ \Ul=\prod_{a=1}^{M}\qe(\w_{2a-1})\;\;
   \Fl\;\prod_{a=1}^{M}\qe(\w_{2a})                       \]
provided $\Fl$ is the `flip' operator
\[ \Fl^{-1}\v^{}_j \Fl=\v^{-1}_j                          \]
while the function $\qe$ solves the following functional
equation:
\[ \frac{\qe(qz)}{\qe(q^{-1}z)}=\frac{1}{1+z}      .      \]
Indeed, let us plug the definition of
the $\v_{j,k}$'s into the hypothetical equation
\bea \Ul^{-b-1}\v_{2a}\Ul^{b+1}\Ul^{-b}\v_{2a}\Ul^{b}-q^{-1}
   \Ul^{-b}\v_{2a-1}\Ul^{b}\Ul^{-b}\v_{2a+1}\Ul^{b}=1   \\\\
   \Ul^{-b}\v_{2a-1}\Ul^{b}\Ul^{-b+1}\v_{2a-1}\Ul^{b-1}
 -q^{-1}\Ul^{-b}\v_{2a-2}\Ul^{b}\Ul^{-b}\v_{2a}\Ul^{b}=1\eea
and dispose of as many $\Ul$'s as possible:
\bea \v_{2a}\Ul\v_{2a}
   -q^{-1}\Ul\v_{2a-1}\v_{2a+1}=\Ul&&\\\\
   \v_{2a-1}\Ul\v_{2a-1}
   -q^{-1}\v_{2a-2}\v_{2a}\Ul=\Ul&&     .               \eea
Then all the $\qe(\w)$ factors but one go the same way which
results in
\bea \v_{2a}\Fl\qe(\w_{2a})\v_{2a}
   -q^{-1}\Fl\qe(\w_{2a})\v_{2a-1}\v_{2a+1}
   &=&\Fl\qe(\w_{2a})                                   \\\\
   \v_{2a-1}\qe(\w_{2a-1})\Fl\v_{2a-1}
   -q^{-1}\v_{2a-2}\v_{2a}\qe(\w_{2a-1})\Fl
   &=&\qe(\w_{2a-1})\Fl    \quad        .               \eea
Once $\Fl$ is gone too, we are left with  
\bea \v^{-1}_{2a}\qe(\w^{}_{2a})\v^{}_{2a}
   -q^{-1}\qe(\w^{}_{2a})\v^{}_{2a-1}\v^{}_{2a+1}
   &=&\qe(\w^{}_{2a})                                   \\\\
   \v^{}_{2a-1}\qe(\w^{}_{2a-1})\v^{-1}_{2a-1}
   -q^{-1}\v^{}_{2a-2}\v^{}_{2a}\qe(\w^{}_{2a-1})
   &=&\qe(\w^{}_{2a-1})                                 \eea
which is nothing but the above functional equation mated
with the commutation relations which closed the last
Section:
\[ \v^{-1}_{2a}\qe(\w^{}_{2a})\v^{}_{2a}
   =\qe(q^{-2}\w^{}_{2a})\qquad\qquad
   \v^{}_{2a-1}\qe(\w^{}_{2a-1})\v^{-1}_{2a-1}
   =\qe(q^{-2}\w^{}_{2a-1})                               \]
\[ \qe(q^{-2}\w_{j})-q^{-1}\qe(\w_{j})\w_{j}
   =\qe(q^{-2}\w_{j})
   -q^{-1}\w_{j}\qe(\w_{j})=\qe(\w_{j})\quad.             \]
So, since everything eventually reduces to that
functional equation, let us see if it can be solved.

\section{q-exponent}

Indeed, the equation in question
\[ \frac{\qe(qz)}{\qe(q^{-1}z)}=\frac{1}{1+z}             \]
is readily fulfilled by those ubiquitous q-exponents
\bea &&\qe(z)=(-qz;q^2)^{}_\infty\qquad
   \mbox{- good for $|q|<1$}                            \\\\
   && \qe(z)=\frac{1}{(-q^{-1}z;q^{-2})^{}_\infty}\qquad
   \mbox{- good for $|q|>1$}                            \eea
where
\[ (x;y)^{}_\infty\equiv\prod_{p=0}^\infty(1-xy^p).       \]
There is no solution entire in $z$ if $|q|=1$.
For those opted for the Weyl-style algebra of observables
(see Section 3) this is the end to the story, not a happy
one in the latter case where the equations of motion survive
but the evolution automorphism behind them turns outer.

The Heisenberg Way has more solutions to its disposal:
\[ \qe_{\mathrm{Heisenberg}}(z)=\qe_{\mathrm{Weyl}}(z)
   \times\mbox{any function}(z^\theta)                    \]
with $\theta$ being the first solution
\[ \theta={\frac{\pi}{\hbar\gamma}}                       \]
to the equation
\[ q^{2\theta}=1                                          \]
with a clear purpose:
\[ \v^\theta_i\w^{}_j=\w^{}_j\v^\theta_i\qquad\qquad
   \v^{}_i\w^\theta_j=\w^\theta_j\v^{}_i  .               \]
Among them we find the one capable of surviving under
$|q|\rightarrow1$ limit (Faddeev 1994):
\[ \qem(z)=\qed(z^\theta)=\left\{\begin{array}{ll}
   \displaystyle\frac{(-qz;q^2)^{}_\infty}
   {(-q^{-\theta^2}z^\theta;q^{-2\theta^2})^{}_\infty}
   \qquad&|q|<1\\\\ \displaystyle
   \frac{(-q^{\theta^2}z^\theta;q^{2\theta^2})^{}_\infty}
   {(-q^{-1}z;q^{-2})^{}_\infty}&|q|>1.\end{array}\right. \]
It is plain to see what makes it so special.
{\em Duality} is the word. We get two functional equations
\[ \frac{\qem(qz)}{\qem(q^{-1}z)}=\frac{1}{1+z}
   \qquad\qquad\frac{\qed(q^{\theta^2}z^\theta)}
   {\qed(q^{-\theta^2}z^\theta)}=\frac{1}{1+z^\theta}     \]
satisfied at once. Consequently, the updated evolution
operator
\[ \Ul=\prod_{a=1}^{M}\qem(\w^{}_{2a-1})\;\;
   \Fl\;\prod_{a=1}^{M}\qem(\w^{}_{2a})
   =\prod_{a=1}^{M}\qed(\w^\theta_{2a-1})\;\;
   \Fl\;\prod_{a=1}^{M}\qed(\w^\theta_{2a})               \]
\[ \Fl^{-1}\v^{}_j \Fl=\v^{-1}_j \qquad\qquad
   \Fl^{-1}\v^{\theta}_j \Fl=\v^{-\theta}_j               \]
not only inherits the right evolution of the pure $\v$'s
\[ \v_{j,k+1}\v_{j,k-1}-q^{-1}\v_{j-1,k}\v_{j+1,k}=1      \]
but also produces an equally right evolution
\[ \v^\theta_{j,k+1}\v^\theta_{j,k-1}
   -q^{-\theta^2}\v^\theta_{j-1,k}\v^\theta_{j+1,k}=1     \]
of their dual twins $\v^\theta$.             
Since the Heisenberg-setup turned out to be just a pair of
decoupled Weyl-ones, we will stick to the latter for the
rest of the paper.

\section{A different angle}

Although the ingredients used in the formula of the
evolution operator are all more or less familiar, they are
put together in a bizarre way. A traditional R-matrix
philosophy would offer a different approach which
we now start presenting. The equation remains the
same
\[ \vc_{m+1,n+1}\vc_{m,n}-q^{-1}\vc_{m,n+1}\vc_{m+1,n}=1  \]
but the change for a kind of `light-cone' variables
\[ j=m-n\qquad\qquad k=m+n                                \]
signals that the  $n$-direction is now considered temporal.
So, we are going to find out what algebra of observables and
what evolution operator produce the evolution
\[ \vc_{m,n+1}=\Uc^{-1}\vc_{m,n}\Uc                       \]
\[ \vc_{m,0}=\vc_m                                        \]
solving that `light-cone' equation.

\section{Algebra of observables ii}

Here is the complete list of relations defining the new
algebra of observables:
\bea && \vc_\ell\vc_m=\vc_m\vc_\ell\qquad
   \mbox{if $m-\ell=0,2,\ldots,M-1$}                    \\\\
   && \vc_\ell\vc_m=q\vc_m\vc_\ell\qquad
   \mbox{if $m=1,3,\ldots,M$}                           \\\\
   && \vc_m\C=q\C\vc_m                                  \\\\
   && \vc_{m+M}=q^\half\C\vc_m\qquad\qquad
   M\equiv1\pmod2  .                                    \eea
This is exactly what it takes to achieve the required
relationship
\bea && \vc_m\wc_m=q^2\wc_m\vc_m                        \\\\
   && \vc_\ell\wc_m=\wc_m\vc_\ell\qquad
   \mbox{if $\ell\neq m\pmod{M}$}                       \eea
between the $\vc$'s and their `derivatives' 
\[ \wc_m=\frac{\vc_{m+1}}{\vc_{m-1}}                      \]
which themselves form the much advertized by the authors
`lattice current algebra'
\bea &&\wc_{m-1}\wc_m=q^2\wc_m\wc_{m-1}                 \\\\
   && \wc_\ell\wc_m=\wc_m\wc_\ell\qquad
   \mbox{if $|m-\ell|\neq 1\pmod{M}$}                   \\\\
   && \wc_{m+M}=\wc_m   .                               \eea
We already met similar relations, it was the end of
Section 3. That time they did not contradict the periodicity
of the $\v$'s. Now they do: it is impossible to have $\C=1$
and good commutation relations at the same time. We shall
soon see why.

\section{Evolution operator ii}

Let us see what the operator
\[ \Ucn=\Sc\Fc\qe(\wc_M)\ldots\qe(\wc_2)\qe(\wc_1)        \]
can do. Of course, the function $\qe$ and the flip operator
$\Fc$ are the same as before
\[ \frac{\qe(qz)}{\qe(q^{-1}z)}=\frac{1}{1+z}             \]
\[ \Fc^{-1}\vc^{}_m\Fc=\vc^{-1}_{m}                       \]
while $\Sc$ is the shift operator
\[ \Sc^{-1}\vc_m\Sc=\vc_{m-1}                             \]
We plug the `naive' evolution into
the hypothetical equation:
\[ \Ucn^{-n-1}\vc_{m+1}\Ucn^{n+1}\Ucn^{-n}\vc_{m}\Ucn^{n}
   \!\!-q^{-1}\Ucn^{-n-1}\vc_{m}\Ucn^{n+1}
   \Ucn^{-n}\vc_{m+1}\Ucn^{n}\!\!=1                       \]
and dispose of as many $\Ucn$'s as possible:
\[ \vc_{m+1}\Ucn\vc_{m}-q^{-1}\vc_{m}\Ucn\vc_{m+1}=\Ucn . \]
Then all the $\qe(\wc)$ factors but one go the same way
which results in
\[ \vc_{m+1}\Sc\Fc\qe(\wc_m)\vc_{m}
   -q^{-1}\vc_{m}\Sc\Fc\qe(\wc_m)\vc_{m+1}
   =\Sc\Fc\qe(\wc_m) .                                    \]
Once $\Sc$ and $\Fc$ are gone too, we are left with
\[ \vc^{-1}_{m}\qe(\wc^{}_m)\vc^{}_{m}-q^{-1}
   \vc^{-1}_{m-1}\qe(\wc^{}_m)\vc^{}_{m+1}=\qe(\wc^{}_m)  \]
which is nothing but our functional equation mated
with the commutation relations which closed the last
Section:
\[ \vc^{-1}_{m}\qe(\wc_m)\vc^{}_{m}=\qe(q^{-2}\wc_m)      \]
\[ \qe(q^{-2}\wc_m)-q^{-1}\qe(\wc_m)\wc_m=\qe(\wc_m)  .   \]
This proves that the `naive' evolution satisfies the
required equations of motion ... as long as $m$ is
neither 1 nor  $M$. We could not reasonably expect it to
do any better because $\Ucn$ obviously had no
respect to the translational symmetry of the algebra of
observables. In order to have this eventually repaired,
let us first figure out how that sad
dependence on the starting point can be cured in a
simpler situation, say, for a monomial 
$\wc^{p_M}_M\ldots\wc^{p_2}_2\wc^{p_1}_1$.
Pulling the factors from the very right to the very left
one by one we get a clear picture of how the ordered
monomials with matching powers but different starting
points turn into each other:
\bea &&\wc^{p_M}_M\ldots\wc^{p_2}_2\wc^{p_1}_1          \\\\
   && =q^{2p_M p_1}q^{-2p_1 p_2}\;\wc^{p_1}_1\wc^{p_M}_M
   \ldots \wc^{p_3}_3\wc^{p_2}_2                        \\\\
   && =q^{2p_M p_1}q^{-2p_2 p_3}\;\wc^{p_2}_2\wc^{p_1}_1
   \ldots \wc^{p_4}_4\wc^{p_4}_3=\ldots                 \eea
Now we know. The expression
$q^{-2p_{m-1} p_m}\;\wc^{p_{m-1}}_{m-1}
   \wc^{p_{m-2}}_{m-2}\ldots\wc^{p_{m+1}}_{m+1}\wc^{p_m}_m$
does not depend on $m$ provided $p_{\ell+M}\equiv p_{\ell}$.
We award it with a self-explanatory notation
\[ \crop\wc^{p_\ell}_\ell\equiv q^{-2p_M p_1}\;
   \wc^{p_M}_M\ldots\wc^{p_2}_2\wc^{p_1}_1                \]
and extend this definition linearly to the corresponding
polynomials, in particular,
\bea &&\crop\qe(\wc_\ell)\equiv\sum_{p_M,\cdots,p_2,p_1}
   c_{p_M}\cdots c_{p_2}c_{p_1}\crop\wc^{p_\ell}        \\\\
   && =\sum_{p_M,\cdots,p_2,p_1}c_{p_M}\cdots c_{p_2}c_{p_1}
   q^{-2p_Mp_1}\;\wc^{p_M}_M\ldots\wc^{p_2}_2\wc^{p_1}_1\\\\
   && =\sum_{p_M,p_1}c_{p_M}c_{p_1}
   q^{-2p_M p_1}\;\wc^{p_M}_M\bigg(\qe(\wc^{}_{M-1})\ldots
   \qe(\wc^{}_3)\qe(\wc^{}_2)\bigg)\wc^{p_1}_1          \\\\
   && =\sum_{p_{m-1},p_m}c_{p_{m-1}}c_{p_m}
   q^{-2p_{m-1} p_m}\;\wc^{p_{m-1}}_{m-1}
   \bigg(\qe(\wc^{}_{m-2})\ldots
   \qe(\wc^{}_{m+2})\qe(\vc^{}_{m+1})\bigg)\wc^{p_m}_m  \eea
with the coefficients $c$ coming from
\[ \qe(z)=\sum_p c_p z^p.                                 \]
The same treatment applies as well to the `selfdual' $\qe$'s
of Section 5:
\[ \qem(z)=\sum_{p,r} c_{p,r} z^p z^{\theta r}            \]
\[ \crop\qem(\wc_\ell)
   =\sum_{p_M,p_1,r_M,r_1}c_{p_M,r_M}c_{p_1,r_1}
   q^{-2(p_M p_1+\theta^2 r_M r_1)}                       \]
\[ \times \wc^{p_M}_M\wc^{\theta r_{M}}_M
   \bigg(\qem(\wc^{}_{M-1})\ldots
   \qem(\wc^{}_3)\qem(\wc^{}_2)\bigg)
   \wc^{p_1}_1\wc^{\theta r_1}_1      .                   \]
So, we achieve the vital translational invariance
of those products
\[ \Sc\crop\qe(\wc_\ell)=\crop\qe(\wc_\ell)\;\Sc          \]
sacrificing none of their `orderness'. 
The time has come to plug the repaired evolution operator
\[ \Uc=\Sc\Fc\crop\qe(\wc_\ell)                           \]
into the hypothetical equation ... see the beginning of
this Section.

Now as we finally established that the operator $\Uc$ is
indeed responsible for the quantized and fully discretized
Liouville equation
\[ \vc_{m+1,n+1}\vc_{m,n}-q^{-1}\vc_{m,n+1}\vc_{m+1,n}=1, \]
we must admit that so far the commitment to this particular
equation was only a matter of personal taste. What would
change if the function involved
\[ \Uc=\Sc\Fc\crop\f(\wc_\ell)                            \]
was different, for instance,
\[ \f(z)=\frac{\qe(z)}{\qe(q^{2\lambda}z)}\qquad?         \]
Nothing except the r.h.s. of the functional equation
\[ \frac{\f(qz)}{\f(q^{-1}z)}=\frac{1+q^{2\lambda}z}{1+z} \]
and the form of the eventual equations of motion
\[ \vc_{m+1,n+1}\vc_{m,n}-q^{-1}\vc_{m,n+1}\vc_{m+1,n}
   =1-q^{\lambda+1}\vc_{m+1,n+1}\vc_{m,n+1}
   \vc_{m+1,n}\vc_{m,n}            .                      \]
By the way, this is another Hirota's equation, the  discrete
sine-Gordon one. We shall see it again in Section 11.

\section{Classical continuum limit}

The equation of Section 6 certainly turns into the
`light-cone' Liouville equation
\[ \frac{\partial^2\psi}{\partial\xi\partial\tau}
   =e^{-2\psi} ,                                          \]
the matching Cauchy problem being
\[ \psi(\xi,\tau)|_{_{\tau=0}}=\psi(\xi)       .          \]
The algebra of observables from Section 7 transforms into
no less familiar Poisson bracket reading
\[ \{\psi(\xi),\psi(\eta)\}
   ={\scriptstyle\frac{1}{4}}\sign(\xi-\eta)    .         \]
The evolution operator of Section 8 bears some resemblance
to the corresponding Hamiltonian
\[ \H=\int d\xi\; e^{-2\psi}          .                   \]
What is wrong? We seem to inherit also the quasiperiodic
boundary condition
\[ \psi(\xi+\pi)=\psi(\xi)+\Psi                           \]
which obviously contradicts to the equation.
The periodic condition
\[ \psi(\xi+\pi)=\psi(\xi)                                \]
could do but that in turn would contradict the Poisson
bracket. A more careful examination reveals that
the `constant' in boundary conditions is not a constant of
motion:
\[ \Uc^{-1}\C\Uc=\C^{-1}           .                      \]
Of course, the lattice equations of motion themselves have
no problem with that. However, their solutions are not
smooth enough to survive a straightforward continuum limit. 
Anyway, this peculiarity is not too relevant to what
we are after in this paper.

\section{Conservation laws}

Looking at the two evolution operators we now possess
\[ \Ul=\prod_{a=1}^{M}\qe(\w_{2a-1})\;\;
   \Fl\;\prod_{a=1}^{M}\qe(\w_{2a})\qquad\qquad
   \Uc=\Sc\Fc\crop\qe(\wc_\ell)                           \]
do we see something in the latter that was not there in the
former? We do, the latter looks almost like a good old
ordered product
of `fundamental R-matrices' (Tarasov {\em et al.}1983). 
According to (Volkov 1997a), the shift-n-flipless part
of $\Uc$
\[ \Omega=\crop\qe(\wc_\ell)                              \]
belongs
\[ \Omega=\Omega\binf                                     \]
in a family
\[ \Omega\bla=\crop\qe(\lambda|\wc_m)\qquad\qquad
   \qe(\lambda|z)\equiv
   \frac{\qe(z)}{\qe(q^{2\lambda} z)}                     \]
consolidated by the Artin-Yang-Baxter's Equation
\[ \R_{m+1}\blamu\R_{m}\bla\R_{m+1}\bmu
   =\R_{m}\bmu\R_{m+1}\bla\R_{m}\blamu         .          \]
The choice of notation
\[ \R_m\bla\equiv\qe(\lambda|\wc_m)\qquad\qquad
   \R_{m}\blamu\equiv\frac{\R_{m}\bla}{\R_{m}\bmu}        \]
is meant to emphasize the R-matrix connection.
Let us recall how that AYBE could be verified. From
(Faddeev and Volkov 1993) comes the multiplication rule
\[ \qe(\lambda|\b)\qe(\lambda|\a)=\qe(\lambda|\a+\b+q\b\a)\]
applying whenever $\a$ and $\b$ satisfy the Weyl's algebra
\[ \a\b=q^2\b\a  .                                        \]
The two $\wc$'s next to each other certainly do, therefore
\[ \R_{m+1}\bla\R_{m}\bla
   =\qe(\lambda|\wc_m+\wc_{m+1}+q\wc_{m+1}\wc_m)  ,       \]
therefore
\bea &&\R_{m+1}\bla\R_{m}\bla\R_{m+1}\bmu\R_{m}\bmu     \\\\
   &&\qquad=\qe(\lambda|\wc_m+\wc_{m+1}+q\wc_{m+1}\wc_m)
   \qe(\mu|\wc_m+\wc_{m+1}+q\wc_{m+1}\wc_m)             \\\\
   &&\qquad=\qe(\mu|\wc_m+\wc_{m+1}+q\wc_{m+1}\wc_m)
   \qe(\lambda|\wc_m+\wc_{m+1}+q\wc_{m+1}\wc_m)         \\\\
   &&=\R_{m+1}\bmu\R_{m}\bmu\R_{m+1}\bla\R_{m}\bla.     \eea
This is it.

AYBE and ordered products are known to make a natural match
\bea && \frac{\R_{m+1}\bla}{\R_{m+1}\bmu}
   \bigg(\R_{m}\bla\R_{m-1}\bla\ldots\R_{1}\bla\bigg)
   \bigg(\R_{m+1}\bmu\R_{m}\bmu\ldots\R_{2}\bmu\bigg)   \\\\
   &&\qquad=\frac{\R_{m+1}\bla}{\R_{m+1}\bmu}
   \bigg(\R_{m}\bla\R_{m+1}\bmu\bigg)
   \bigg(\R_{m-1}\bla\R_{m}\bmu\bigg)\ldots
   \bigg(\R_{1}\bla\R_{2}\bmu\bigg)                     \\\\
   && \qquad=\bigg(\R_{m}\bmu\R_{m+1}\bla\bigg)
   \frac{\R_{m}\bla}{\R_{m}\bmu}
   \bigg(\R_{m-1}\bla\R_{m}\bmu\bigg)\ldots
   \bigg(\R_{1}\bla\R_{2}\bmu\bigg)                     \\\\
   && \qquad=\bigg(\R_{m}\bmu\R_{m+1}\bla\bigg)
   \bigg(\R_{m-1}\bmu\R_{m}\bla\bigg)
   \frac{\R_{m-1}\bla}{\R_{m-1}\bmu}\ldots
   \bigg(\R_{1}\bla\R_{2}\bmu\bigg)                     \\\\
   && \qquad=\bigg(\R_{m}\bmu\R_{m+1}\bla\bigg)
   \bigg(\R_{m-1}\bmu\R_{m}\bla\bigg)\ldots
   \frac{\R_{2}\bla}{\R_{2}\bmu}
   \bigg(\R_{1}\bla\R_{2}\bmu\bigg)                     \\\\
   && \qquad=\bigg(\R_{m}\bmu\R_{m+1}\bla\bigg)
   \bigg(\R_{m-1}\bmu\R_{m}\bla\bigg)\ldots
   \bigg(\R_{1}\bmu\R_{2}\bla\bigg)\R_{1}\blamu         \\\\
   &&=\bigg(\R_{m}\bmu\R_{m-1}\bmu\ldots\R_{1}\bmu\bigg)
   \bigg(\R_{m+1}\bla\R_{m}\bla\ldots\R_{2}\bla\bigg)
   \frac{\R_{1}\bla}{\R_{1}\bmu} ,                      \eea
so, one hopes the $\circlearrowright$-ed product to step
beyond $m=M-2$ and deliver
\[ \Omega\bla\Omega\bmu=\Omega\bmu\Omega\bla    .         \]
Let us see.
\bea &&\Omega\bla \Omega\bmu                            \\\\
   &&\qquad=\sum c_{p_M}^\sla  c_{p_1}^\sla q^{-2 p_M p_1}
   \wc^{p_M}_M\R_{\!M\!-\!1}^\sla  \ldots
   \R_2^\sla  \wc^{p_1}_1                               \\\\
   &&\qquad\times\sum 
   c_{r_M}^\smu c_{r_1}^\smu q^{-2 r_M r_1} 
   \wc^{r_M}_M\R_{\!M\!-\!1}^\smu  \ldots
   \R_2^\smu  \wc^{r_1}_1                               \\\\
   &&\qquad=\sum c_{p_M}^\sla  c_{r_M}^\smu  
   c_{p_1}^\sla  c_{r_1}^\smu 
   q^{-2(p_M p_1+r_M r_1+r_M p_1)}                      \\\\
   &&\qquad\times\wc^{p_M}_M
   \R_{\!M\!-\!1}^\sla  \wc^{r_M}_M
   \R_{\!M-2}^\sla  \R_{\!M\!-\!1}^\smu  \ldots
   \R_2^\sla  \R_3^\smu  
   \wc^{p_1}_1\R_2^\smu  
   \wc^{r_1}_1                                          \eea
- so far, we only recalled the definitions and did some
reshuffling not involving any nontrivial commutation
relations, $\bla$ and $\bmu$ moved to the superscript level
in order to save some paper -
\bea && =\sum c_{p_M}^\sla  c_{r_M}^\smu  
   c_{p_1}^\sla  c_{r_1}^\smu 
   q^{-2(p_M p_1+r_M r_1+r_M p_1)}                      \\\\
   &&\times\wc^{p_M}_M
   \R_{\!M\!-\!1}^\sla  \wc^{r_M}_M
   \frac{\R_{\!M\!-\!1}^\smu }{\R_{\!M\!-\!1}^\sla }
   \frac{\R_{\!M\!-\!1}^\sla }{\R_{\!M\!-\!1}^\smu }
   \R_{\!M-2}^\sla  \R_{\!M\!-\!1}^\smu \ldots
   \R_2^\sla  \R_3^\smu  
   \wc^{p_1}_1\R_2^\smu  
   \wc^{r_1}_1                                          \eea
- the unit operator $\frac{\R_{\!M\!-\!1}^\smu }
{\R_{\!M\!-\!1}^\sla }\frac{\R_{\!M\!-\!1}^\sla }
{\R_{\!M\!-\!1}^\smu }=\mathsf{1}$ has been inserted -
\bea  &&=\sum c_{p_M}^\sla  c_{r_M}^\smu  
   c_{p_1}^\sla  c_{r_1}^\smu 
   q^{-2(p_M p_1+r_M r_1+r_M p_1)}                      \\\\
   &&\times\wc^{p_M}_M
   \R_{\!M\!-\!1}^\sla  \wc^{r_M}_M
   \frac{\R_{\!M\!-\!1}^\smu }{\R_{\!M\!-\!1}^\sla }
   \R_{\!M-2}^\smu  \R_{\!M\!-\!1}^\sla  \ldots
   \R_2^\smu  \R_3^\sla  \frac{\R_2^\sla }{\R_2^\smu }
   \wc^{p_1}_1\R_2^\smu  
   \wc^{r_1}_1                                          \eea
- AYBE did its habitual job -
\bea &&=\sum 
   c_{p_M}^\sla  c_{r_M}^\smu  
   c_{p_1}^\sla  c_{r_1}^\smu 
   q^{-2(p_M p_1+r_M r_1+r_M p_1)}                      \\\\
   &&\times\wc^{p_M}_M
   \R_{\!M\!-\!1}^\sla  \wc^{r_M}_M
   \frac{\R_{\!M\!-\!1}^\smu }{\R_{\!M\!-\!1}^\sla }\ldots
   \R_{m+1}^\smu  \R_{m+2}^\sla  
   \R_{m}^\smu \R_{m+1}^\sla 
   \R_{m-1}^\smu  \R_{m}^\sla  \ldots
   \frac{\R_2^\sla }{\R_2^\smu }
   \wc^{p_1}_1\R_2^\smu  
   \wc^{r_1}_1                                          \eea
- nothing happened, we just refocused the attention to the
middle portion of the product -
\bea &&=\sum c_{p_M}^\sla  c_{r_M}^\smu  
   c_{r_{m+1}}^\smu  c_{r_{m}}^\smu  
   c_{p_{m+1}}^\sla  c_{p_{m}}^\sla  
   c_{p_1}^\sla  c_{r_1}^\smu 
   q^{-2(p_M p_1+r_M r_1+r_M p_1)}                      \\\\
   &&\times\wc^{p_M}_M
   \R_{\!M\!-\!1}^\sla  \wc^{r_M}_M
   \frac{\R_{\!M\!-\!1}^\smu }{\R_{\!M\!-\!1}^\sla }\ldots
   \wc^{r_{m+1}}_{m+1}
   \R_{m+2}^\sla  
   \wc^{r_{m}}_{m}\wc^{p_{m+1}}_{m+1}
   \R_{m-1}^\smu  
   \wc^{p_{m}}_{m}\ldots
   \frac{\R_2^\sla }{\R_2^\smu }
   \wc^{p_1}_1\R_2^\smu  
   \wc^{r_1}_1                                          \eea
- we disassembled some of the $\R$'s -
\bea &&=\sum c_{p_M}^\sla  c_{r_M}^\smu  
   c_{r_{m+1}}^\smu  c_{p_{m+1}}^\sla 
   c_{r_{m}}^\smu  c_{p_{m}}^\sla  
   c_{p_1}^\sla  c_{r_1}^\smu 
   q^{-2(p_M p_1+r_M r_1+r_M p_1-p_{m+1}r_m)}           \\\\
   &&\times\bigg(\wc^{p_M}_M
   \R_{\!M\!-\!1}^\sla  \wc^{r_M}_M
   \frac{\R_{\!M\!-\!1}^\smu }{\R_{\!M\!-\!1}^\sla }\ldots
   \wc^{r_{m+1}}_{m+1}\R_{m+2}^\sla 
   \wc^{p_{m+1}}_{m+1}\bigg)                            \\\\
   &&\times\bigg(\wc^{r_{m}}_{m}
   \R_{m-1}^\smu \wc^{p_{m}}_{m}\ldots
   \frac{\R_2^\sla }{\R_2^\smu }
   \wc^{p_1}_1\R_2^\smu \wc^{r_1}_1\bigg)               \eea
- the two in the middle traded places -
\bea &&=\sum c_{r_{m}}^\smu  c_{p_{m}}^\sla  
   c_{p_1}^\sla  c_{r_1}^\smu 
   c_{p_M}^\sla  c_{r_M}^\smu  
   c_{r_{m+1}}^\smu  c_{p_{m+1}}^\sla 
   q^{-2(r_m r_{m+1}+p_m p_{m+1}+p_m r_{m+1}-r_1p_M)}   \\\\
   &&\times\bigg(\wc^{r_{m}}_{m}
   \R_{m-1}^\smu \wc^{p_{m}}_{m}\ldots
   \frac{\R_2^\sla }{\R_2^\smu }
   \wc^{p_1}_1\R_2^\smu \wc^{r_1}_1\bigg)              \\\\
   &&\times\bigg(\wc^{p_M}_M
   \R_{\!M\!-\!1}^\sla  \wc^{r_M}_M
   \frac{\R_{\!M\!-\!1}^\smu }{\R_{\!M\!-\!1}^\sla }\ldots
   \wc^{r_{m+1}}_{m+1}\R_{m+2}^\sla 
   \wc^{p_{m+1}}_{m+1}\bigg)                            \eea
- the two halves in big brackets passed through each other -
\bea &&=\sum c_{r_{m}}^\smu  c_{p_{m}}^\sla  
   c_{p_1}^\sla  c_{p_M}^\sla 
   c_{r_1}^\smu  c_{r_M}^\smu  
   c_{r_{m+1}}^\smu  c_{p_{m+1}}^\sla 
   q^{-2(r_m r_{m+1}+p_m p_{m+1}+p_m r_{m+1})}          \\\\
   &&\times\wc^{r_{m}}_{m}
   \R_{m-1}^\smu \wc^{p_{m}}_{m}\ldots
   \frac{\R_2^\sla }{\R_2^\smu }
   \wc^{p_1}_1\R_2^\smu 
   \wc^{p_M}_M\wc^{r_1}_1
   \R_{\!M\!-\!1}^\sla  \wc^{r_M}_M
   \frac{\R_{\!M\!-\!1}^\smu }{\R_{\!M\!-\!1}^\sla }\ldots
   \wc^{r_{m+1}}_{m+1}\R_{m+2}^\sla 
   \wc^{p_{m+1}}_{m+1}                                  \eea
- the two in the middle traded places -
\bea &&=\sum c_{r_{m}}^\smu  c_{p_{m}}^\sla  
   c_{r_{m+1}}^\smu  c_{p_{m+1}}^\sla 
   q^{-2(r_m r_{m+1}+p_m p_{m+1}+p_m r_{m+1})}          \\\\
   &&\times\wc^{r_{m}}_{m}
   \R_{m-1}^\smu \wc^{p_{m}}_{m}\!\ldots\!
   \frac{\R_2^\sla }{\R_2^\smu }
   \R_1^\sla \R_2^\smu 
   \R_M^\sla \R_1^\smu 
   \R_{\!M\!-\!1}^\sla  \R_M^\smu 
   \frac{\R_{\!M\!-\!1}^\smu }{\R_{\!M\!-\!1}^\sla }
   \!\ldots\!\wc^{r_{m+1}}_{m+1}\R_{m+2}^\sla 
   \wc^{p_{m+1}}_{m+1}                                  \eea
- we assembled some $\R$'s, now it only remains
to apply AYBE three more times~-
\bea &&\qquad=\sum c_{r_{m}}^\smu  c_{p_{m}}^\sla  
   c_{r_{m+1}}^\smu  c_{p_{m+1}}^\sla 
   q^{-2(r_m r_{m+1}+p_m p_{m+1}+p_m r_{m+1})}          \\\\
   &&\qquad\times\wc^{r_{m}}_{m}
   \R_{m-1}^\smu \wc^{p_{m}}_{m}\ldots
   \R_1^\smu \R_2^\sla 
   \R_M^\smu \R_1^\sla 
   \R_{\!M\!-\!1}^\smu \R_M^\smu \ldots
   \wc^{r_{m+1}}_{m+1}\R_{m+2}^\sla 
   \wc^{p_{m+1}}_{m+1}                                  \\\\
   &&\qquad=\sum c_{r_m}^\smu  c_{r_{m+1}}^\smu 
   q^{-2 r_m r_{m+1}} 
   \wc^{r_m}_m\R_{m-1}^\smu  \ldots
   \R_{m+2}^\smu  \wc^{r_{m+1}}_{m+1}                  \\\\
   &&\qquad\times\sum c_{p_m}^\sla  c_{p_{m+1}}^\sla 
   q^{-2 p_m p_{m+1}} 
   \wc^{p_m}_m\R_{m-1}^\sla  \ldots
   \R_{m+2}^\sla \wc^{p_{m+1}}_{m+1}                    \\\\
   &&=\Omega\bmu \Omega\bla            .                \eea
Done, at least for $M>5$. In fact, even $M=3$ is possible
but this would take three more pages to verify.
Anyway, a more civilized edition of the above proof is
presented in (Volkov 1997b).

The commutativity of the $\Omega$'s may be a good news but
there is a bad one too. The flip operator $\Fc$ does not
commute with the $\Omega$'s. Which means there is another
family
\[ \mho\bla=\Fc^{-1}\Omega\bla\Fc=\Fc\Omega\bla\Fc^{-1}   \]
not coinciding with the original one.
Of course, the $\mho$'s commute with each other
\[ \mho\bla\mho\bmu=\mho\bmu\mho\bla                      \]
but it is not immediately clear whether 
\[ \Omega\bla\mho\bmu=\mho\bmu\Omega\bla                  \]
should also be true. Fortunately, there is some hidden
agenda making it happen. Technically-wise, the proof is the
same as that above except the AYBE in use is somewhat
different:
\bea && \bigg(\wc_{m+1}^{\mu}
   \qe(\lambda\!-\!\mu|\wc_{m+1}^{})\bigg)
   \qe(\lambda|\wc_{m}^{})\qe(\mu|\wc_{m+1}^{-1})       \\\\
   && \qquad=\qe(\lambda\!-\!\mu|\wc_{m+1}^{})
   \qe(\lambda|q^{-2\mu}\wc_{m}^{})
   \wc_{m+1}^{\mu}\qe(\mu|\wc_{m+1}^{-1})               \\\\
   && \qquad=\frac{\qe(\lambda|q^{-2\mu}\wc_{m+1}^{})}
   {\qe(\mu|q^{-2\mu}\wc_{m+1}^{})}
   \qe(\lambda|q^{-2\mu}\wc_{m}^{})
   q^{\mu^2}\qe(\mu|q^{-2\mu}\wc_{m+1}^{})              \\\\
   && \qquad=q^{\mu^2}\qe(\mu|q^{-2\mu}\wc_{m}^{})
   \qe(\lambda|q^{-2\mu}\wc_{m+1}^{})
   \frac{\qe(\lambda|q^{-2\mu}\wc_{m}^{})}
   {\qe(\mu|q^{-2\mu}\wc_{m}^{})}                       \\\\
   && \qquad=\wc_{m}^{\mu}\qe(\mu|\wc_{m}^{-1})
   \qe(\lambda|q^{-2\mu}\wc_{m+1}^{})
   \qe(\lambda\!-\!\mu|\wc_{m}^{})                      \\\\
   && =\qe(\mu|\wc_{m}^{-1})\qe(\lambda|\wc_{m+1}^{})
   \bigg(\qe(\lambda\!-\!\mu|\wc_{m}^{})
   \wc_{m}^{\mu}\bigg)   .                              \eea
Once the flip-n-shift join
\[ \Uc\bla=\Sc\Fc\Omega\bla=\mho\bla\Sc\Fc                \]
one realizes that full commutativity is not there
\[ \Uc\bla\Uc\bmu\neq\Uc\bmu\Uc\bla ,                     \]
only `squares' can do:
\bea && \bigg(\Uc\bka\Uc\bla\bigg)\bigg(\Uc\bmu\Uc\bnu\bigg)
   =\Fc\Omega\bka\Fc\Omega\bla\Fc\Omega\bmu\Fc\Omega\bnu\\\\
   && =\Fc^4\mho\bka\Omega\bla\mho\bmu\Omega\bnu
   =\Fc^4\mho\bmu\Omega\bnu\mho\bka\Omega\bla           \\\\
   && =\Fc\Omega\bmu\Fc\Omega\bnu\Fc\Omega\bka\Fc\Omega\bla=
   \bigg(\Uc\bmu\Uc\bnu\bigg)\bigg(\Uc\bka\Uc\bla\bigg).\eea
In particular,
\[ \Uc^{-2}\Uc^2\bla\Uc^2=\Uc^2\bla    .                  \]
On these grounds, let us call the $\Uc\bla$'s `conservation
laws' even though what actually happens is that only their
squares are only recovered on every other step in time.
This peculiarity has everything to do with that discussed
in Section 9. Apparently, not one but two time steps
should make a `physical' time unit.

\section{Conclusion}

We developed here the scheme allowing to describe
some quantum dynamical systems in discrete 1+1-dimensional
space-time. The discretized Liouville model was taken
as the main example and treated both for laboratory
coordinates and light-like ones. All considerations
were purely algebraic, no representation and/or
Hilbert space was used. We confined ourselves to pure
Heisenberg picture of quantum theory.

The main outcome is the construction of evolution operator
realizing the automorphism of the algebra of observables
leading to the Heisenberg equations of motion representing
lattice and quantum deformation of the corresponding
classical equations. In this construction the famous
q-exponent (q-dilogarithm) played the most prominent part.

We discussed also the integrability of the model presenting
the set of conservation laws. Their construction and the
verification of commutativity was based on a new solution
of Artin-Yang-Baxter relation, being a close relative
of the q-exponent.

We hope that the scheme of this paper is general enough
and allows to include many related models of quantum
theory. Our papers mentioned in Introduction give some
illustration of this.

\end{document}